25·29 August 2024

# inter.noise
NANTES FRANCE

# Denoising by neural network for muzzle blast detection

Hadrien Pujol[1]
Acoem
200 Chemin des Ormeaux, 69760 Limonest, France

Matteo Bevillacqua
Acoem
200 Chemin des Ormeaux, 69760 Limonest, France

Christophe Thirard
Acoem
200 Chemin des Ormeaux, 69760 Limonest, France

Thierry Mazoyer[2]
Acoem
200 Chemin des Ormeaux, 69760 Limonest, France

## 1. ABSTRACT

   *Acoem develops gunshot detection systems, consisting of a microphone array and software that detects and locates shooters on the battlefield.*

   *The performance of such systems is obviously affected by the acoustic environment in which they are operating: in particular, when mounted on a moving military vehicle, the presence of noise reduces the detection performance of the software. To limit the influence of the acoustic environment, a neural network has been developed. Instead of using a heavy convolutional neural network, a lightweight neural network architecture was chosen to limit the computational resources required to embed the algorithm on as many hardware platforms as possible.*

   *Thanks to the combination of a two hidden layer perceptron and appropriate signal processing techniques, the detection rate of impulsive muzzle blast waveforms (the wave coming from the detonation and indicating the position of the shooter) is significantly increased. With a rms value of noise of the same order as the muzzle blast peak amplitude, the detect rate is more than doubled with this denoising processing.*

_______________________

[1] hadrien.pujol@acoem.com
[2] thierry.mazoyer@acoem.com

## 1. INTRODUCTION

In the context of civilian and military protection, ACOEM and its subsidiaries are developing acoustic devices to detect gunshots. One of the challenges is to find the shooter's position from the acoustic information contained in the recorded sound signals. [1, 3]

When the threat, in a military context, is a weapon fire with supersonic munitions, two acoustic impulses (or "events") may be identified. In chronological order: the Mach Shock Wave, produced by a projectile with a velocity greater than the sound speed, and the muzzle blast generated by the shot detonation. The processing of these two events is usually sufficient to obtain with good accuracy the location in distance and direction of the shooter. Other information such as miss distance from the microphone array, and caliber can also be obtained with varying degrees of confidence.

However, when the detector is installed on a vehicle whose engine and other sources generate significant background noise, the detection rate of the muzzle blasts often decreases significantly. At the same time, the quality of the information that can be extracted from these waves deteriorates and the number of false alarms increases, making it more frequent for the shooter to be located incorrectly.

This work therefore proposes a method for overcoming military vehicle noise in muzzle blast processing. Before presenting the results, the operation of the detection system is described. The neural network developed is then presented, along with the specific features used to optimize it. Finally, results are presented and discussed in terms of performance gains for the system.

## 2. BACKGROUND AND ISSUES

This section presents the prerequisites about both the physics of gunfire and the Acoem's system, necessary for a good understanding of the problem.

### 2.1. Presentation of the Acoem's detection shot system

The system is designed to detect supersonic gunshots (so called sniper attack). Two acoustic waves are generated and the system is designed to analyse them. Figure 1 shows the acoustic characteristics of such a threat.

Firstly, the supersonic passage of the projectile generates a shock wave, known as the Mach Wave (MW) [2]. This particularly energetic and high frequency (outside most of common noises) acoustic pressure enables the system to detect a threat. The threat is then fully characterized by detecting the Muzzle Blast (MB) which is generated by the detonation at the moment of firing. Finally, information extracted from the two elements are combined (data fusion) to estimate all characteristics of the threat, by estimating the caliber of the bullet and the shooter's position (distance and angle).

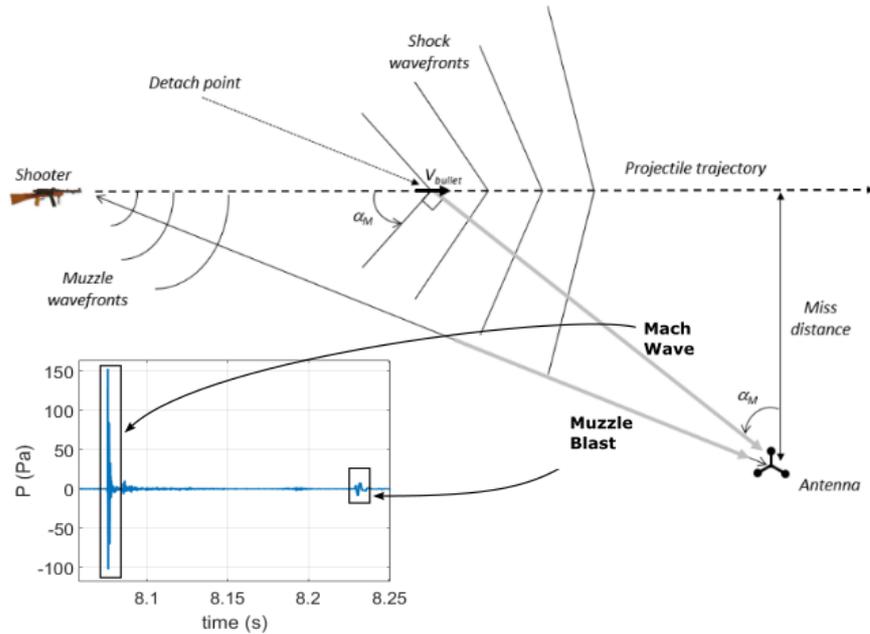

Figure 1: Description of the waves generated by a shot with a supersonic projectile: distance 150 m, miss distance 20 m and caliber 7.62 mm

Figure 1 also shows a characteristic signal recorded by the system in the case of an aggressive shot. There are two fundamental qualitative differences between MW and MB. The first is more energetic (over a hundred Pa versus a dozen), and has a shorter time support (a few ms versus a few tens of ms). This justifies the choice of these MW for threat detection.

## 2.2. Problems related to the use of acoustic system mounted on vehicles

Because the detection of the threat is based on acoustic detection, the performances of the system depend on the ambient noise level. These performances may decrease when the signal to noise ratio (SNR defined hereafter) become more unfavorable. This is particularly the case when the microphone array is deployed on military vehicles, which are known to be noisy. In the presence of vehicle noise, the MW remains easily detectable, as its sound level is always higher than that of the ambient noise (see Figure 1). On the other hand, as shown in Figure 2, when military vehicle noise is added, it is no longer as easy to recognize the acoustic signature of the MB.

However, the detection system of the MB works on the basis of a pulse detection system, which looks for sharp transient changes in the signal. If the MB is of low amplitude compared with the noise, it is indistinguishable from the ambient noise, and may therefore go undetected.

Another disadvantage of the presence of noise in the signal is that the detection will send back many more pulses that can later be interpreted as MB. We therefore understand what is at stake in this work: MB detection is absolutely essential to characterize a gunshot correctly. Since detection performance is degraded in the presence of military vehicle noise, we need to find a way of eliminating this noise or reducing its influence on MB detection performance.

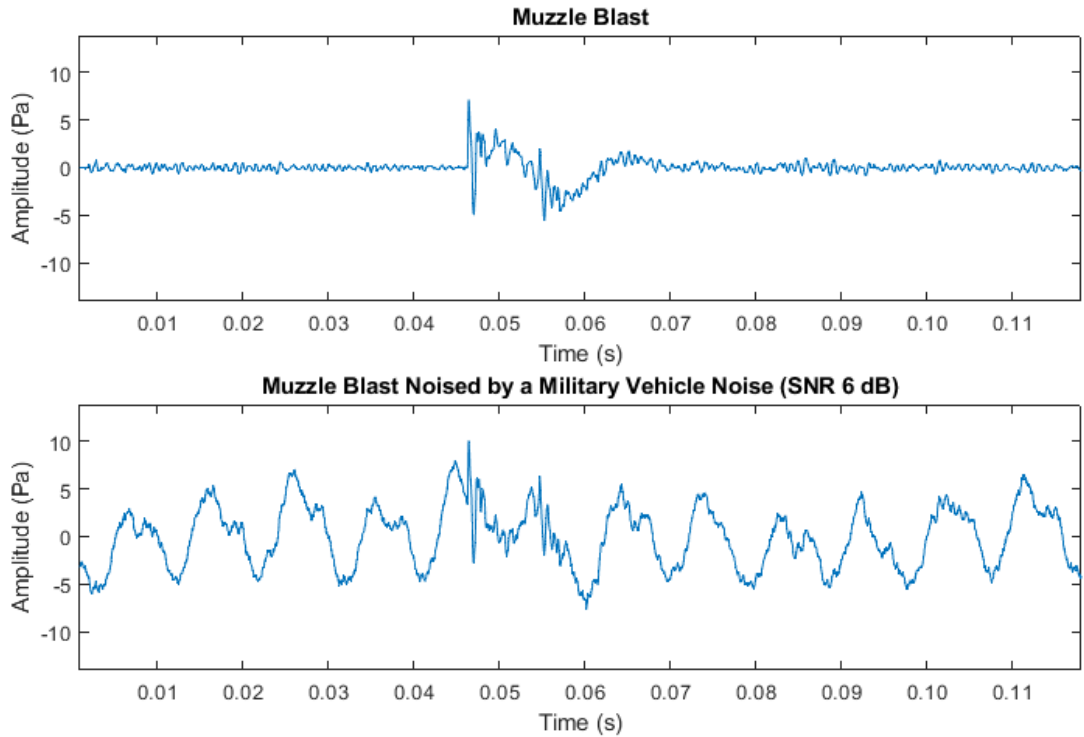

Figure 2 : Example of a noised muzzle blast

## 3. DEVELOPMENT OF A NEURAL NETWORK TO IMPROVE MB DETECTION

From Wiener filters [5] to Wavenet [4], many algorithms have been developed for acoustic denoising. To limit the computational load on the digital signal processor (DSP) of the system, the original approach of this study is neither to use traditional signal processing tools, heavy (conventional) neural networks (NN), but to use a two hidden layer perceptron, surrounded by conventional signal processing.

This section first presents the data used for learning, then the proposed denoising architecture, and finally describes a training technique to promote model convergence in the face of SNR diversity.

### 3.1. Selected dataset

Acquiring the acoustic signature of gunshots requires conditions (security,…) that make measurement campaigns expensive. For this reason, we opted for data augmentation using records of shots fired in a quiet environment (accumulated over several decades of trials by ACOEM) and recordings of vehicle noises. For learning purposes, we chose to work with 800 shots from the same caliber, but in different situations (distance, environment, weapon, etc.). These shots are then noised, by adding signal from one of three different noise recordings of the same military vehicle and the same vehicle speed. This approach provides both the noisy signal from the shot, used as input to the perceptron, and the original, noise-free signal, used as target for training.

To separate the data into train and validation sets, the approach is as follows. The shots are divided into two subsets. Each of the three sounds are subdivided in several sections and also forms a separate subset. Each noise subset is then combined with each shot subset, giving us six noised shot subsets, as shown in Figure 3. A training base is then made up of a MB sub-base noised by two different noises. The validation sub-base then consists of shots from the unused MB sub-base, combined noise not chosen for training. In this way, the shots and noise used for the validation database are totally separate from those used for training. This separation also facilitates implementation of cross-validation process to avoid overfitting while learning.

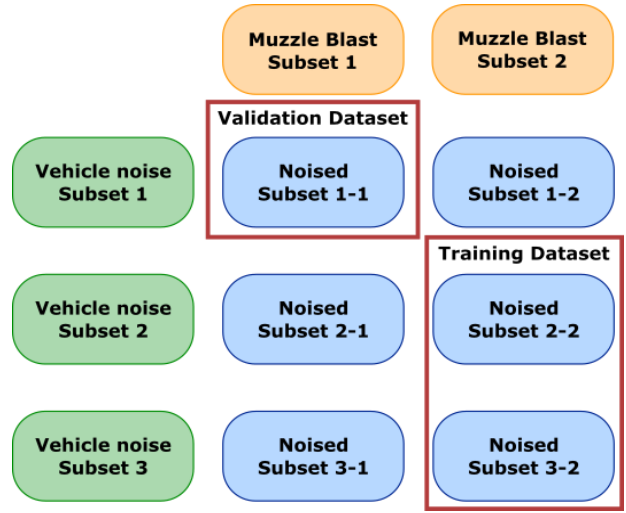

Figure 3 : Distribution of subsets

## 3.2. Neural Network architecture

A large number of neural network architectures have been proposed in the literature for signal denoising [7-12]. Auto-encoders have the advantage of being able to work directly on a raw signal, at the input and output of the network. This explains the choice of this type of architecture.

However, the objective of the present work is to fit in the existing hardware platform which leads to some technological constraints. In particular, the DSP based computing board is not designed for large-scale convolution calculations (no dedicated processing unit like GPUs or NPUs). Given the time required for certification in the military industry, a change of component is not an option. Computation sizing is therefore essential to reconcile performance and real-time calculations.

The system works on time frames of 2,048 samples at a sampling frequency *Fs*. To limit the number of calculations made by the DSP, the signal is down-sampled by a factor of 8, as explained in Figure 4. To avoid aliasing, the signal is also low-pass filtered before decimation. It is important to notice that MB has no energy above *Fs*/16, so this filtering only slightly modifies the shape of the MB. Finally, interpolation is performed to obtain a signal with a length of 2,048 samples, as was the initial signal.

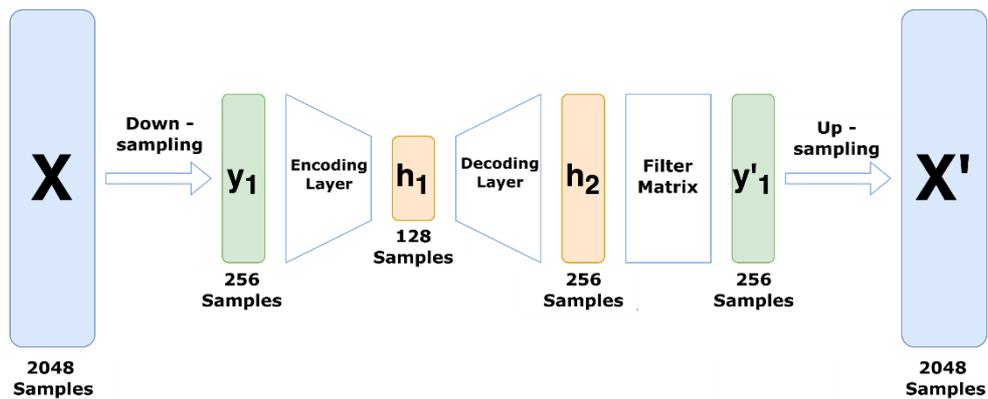

Figure 4 : Architecture of the proposed neural network

The proposed network therefore consists of an autoencoder followed by a layer simulating filtering (cf. Figure 4). The autoencoder is a single hidden layer linking input and output via fully connected connections. The third layer acts as a low-pass filter to avoid high-frequency signal variations due to the encoder's non-linear activation function. This layer matrix is initialized as a low-pass filter (8$^{th}$-order Butterworth filter with a cut-off frequency of *Fs/41*). Indeed, the main idea is to see the filtering operation as a matrix multiplication: Equation 1 describes this mathematically. We assume that this filter has a convolution kernel h and an impulse response r (h(t) = r(-t)). The filter matrix is therefore a square matrix of dimension 256 (the size of our sub-sampled signal), with row n consisting of 0 except for the elements around column n, which are exactly the kernel h.

$$(h * x)(n) \qquad = \sum_{m=1}^{256} h(m)\, x(n - m)$$

$$h * x \quad = \begin{pmatrix} h & \cdots & & 0 \\ & \ddots & & \\ \vdots & & h & \vdots \\ & & & \ddots & \\ 0 & \cdots & & h \end{pmatrix} \cdot \begin{pmatrix} x(1) \\ \cdots \\ x(256) \end{pmatrix} \qquad (1)$$

The strategy used to optimize this filter matrix is detailed in section 3.3.

### 3.3.   Iterative learning

Denoising difficulty increases with the Signal to noise ratio (SNR). In our case, since noise signals are highly impulsive, comparing the effective pressure of MBs (RMS in the whole time window) and vehicle noise is not relevant. In this work, the SNR expressed in dB, is the ratio over the maximum pressure of the MB ($P_{max\_MB}$) and the effective pressure of the vehicle noise ($P_{eff\_noise}$). The SNR is thus calculated according to the formula:

$$SNR = 20 \log \left( \frac{P_{max,MB}}{P_{eff,noise}} \right) \qquad (2)$$

So, to help the algorithm to converge, training is carried out in several phases. Each phase corresponds to the addition of a certain number of examples based on their SNR. Initially, only signals with SNR > 0 dB are selected. In the next phase, all noisy signals with SNR > -5 dB are used, and so on, until -20 dB. In this way, each phase can be seen as transfer learning from the previous phase.

In addition, at the start of each phase, the matrix representing the low-pass filtering at the output of the autoencoder (see Section 3.2 & Figure 4) is frozen. In this way, only the autoencoder coefficients are optimized. Then, after a certain number of iterations, this filter matrix is released, and its coefficients are optimized along with those of the network. Figure 5 shows the various steps of this learning process.

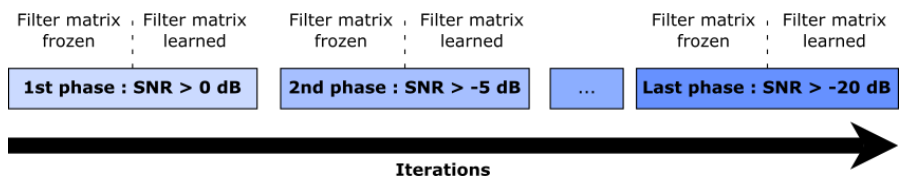

Figure 5 : Learning process

In our case, the metric minimized by the learning is a simple MSE: the difference between the noisy signal *X* and the target (noiseless) signal *Y*. The error E is thus defined by the sum of the squared errors of each temporal sample $X_i$ and $Y_i$ :

$$E = \sum_i (X_i - Y_i)^2 \qquad (3)$$

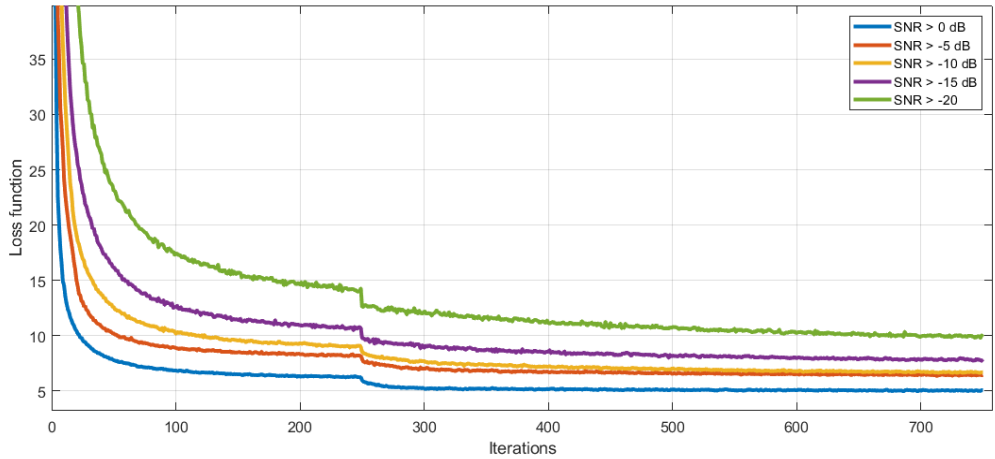

Figure 6 : Convergence curves for each training phase

Figure 5Figure 6 shows the training convergence curves along steps, computed using cross validation. As expected, the convergence curves show two modes of convergence. The first, up to 250 iterations, corresponds to autoencoder optimization, following the addition of new examples with more unfavorable SNR. In this case, we have a high MSE at the beginning of this phase, due to the introduction of the new samples, and rapid convergence following the improvement of the autoencoder. It is also natural to observe that the noisier the data, the slower the convergence, and the higher the convergence asymptote. After 250 iterations, the autoencoder's output filter matrix begins to be optimized. We then see a second rapid convergence, proving the value of optimizing this last layer. We also note that the more advanced the learning phase, the greater the contribution of this last filter matrix.

## 4. RESULTS

Once training is complete, the performance of the network is evaluated within the real system. Denoising results are presented in this section on shots of the same caliber class as the one used for training, then on shots of a different caliber.

### 4.1. Performance criteria

The objective criterion of the MSE between the noise-free signal and the denoised signal is relevant for learning, as it allows us to use a scalar to characterize denoising performance. However, this criterion is not relevant for finely characterizing the denoising performance.

Here, the denoising quality obtained by the network is evaluated in terms of the ability of all surrounding algorithmic blocks of the whole existing system to detect an MB in the midst of vehicle noise. Indeed, beyond the quality of the reconstruction of the noiseless signal, the objective is to know whether this reconstruction retains enough information to fit well in the system to consolidate the detection of the MW with the corresponding MB (see section 2.1).

To evaluate the MB detection, a complete signal (see Figure 1) is presented to the system. The system then automatically processes the entire signal and outputs a number of MB detections. Bearing in mind that for all signals, the start of the MB is known a priori, we'll assume here that a MB is detected if a MB alert is raised, and that the MB detection time corresponds to the true MB time of the noiseless signal, with a slight tolerance set at around Fs/100, where Fs is the sampling frequency. This tolerance value is such that we are, from our knowledge of the whole system behaviour, below the threshold of sensivity.

## 4.2. Results on the concatenation of validation databases

The results of the validation databases are presented here. As explained in section 3.1 , the validation database is made up of shots not seen during training, noised with a recording of a military vehicle not used during training.

For greater statistical precision, and to better qualify the method, all permutations are made so that each of the six sub-bases can be used as a validation base. The results are therefore a concatenation of the performances of six networks, each integrated separately into the system, and validated on its own validation sub-base.

Figure 7 shows the detection rate as a function of SNR. For each point on the curve, a margin of error is indicated by a segment. This margin of error [6] is calculated by assuming that the system's detection rate corresponds to the probability that a MB event will be detected. Let P be this probability and N the number of realizations used to estimate this probability, then the margin $\delta P$ is:

$$\delta P = \sqrt{\frac{P(1-P)}{N}} \tag{4}$$

There is although a 0.68 probability that the true detection rate lies between $P\text{-}\delta P$ and $P\text{+}\delta P$ and a 0.95 probability that it lies between $P\text{-}2\delta P$ and $P\text{+}2\delta P$.

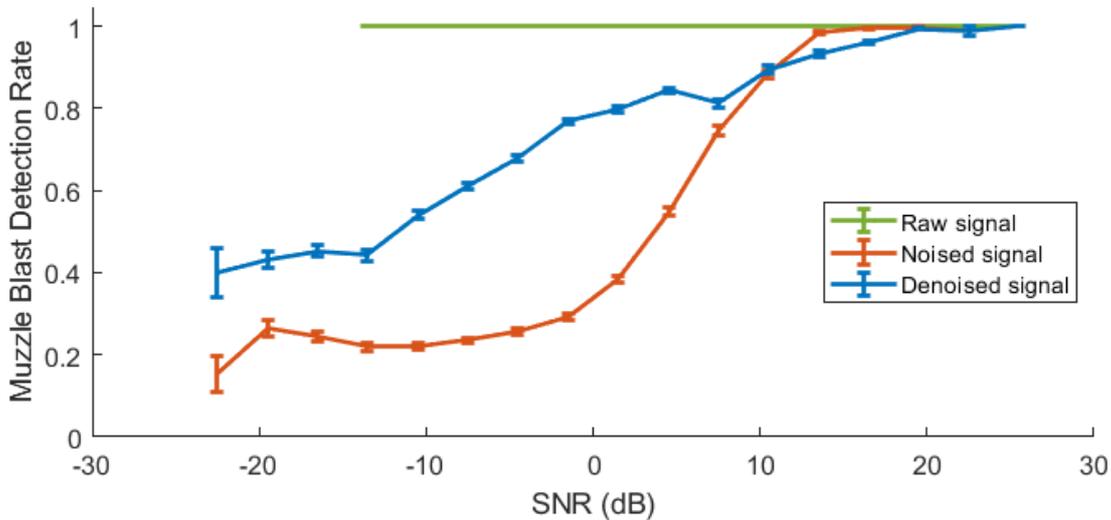

Figure 7 : Improved MB detection rate thanks to denoising on validation datasets

Figure 7 compares the detection performance of MB in a quiet environment (green curve), with that obtained with the same MB superimposed on the noise of a moving military vehicle (orange curve), and finally, in blue, with the same shots after denoising by the proposed system (see section 3.2). From 10 dB SNR upwards, the proposed system offers a significant contribution. In particular, at -2 dB SNR, the detection rate rises from 0.30 to 0.75. MBs are therefore detected more than twice as often. Even at very unfavorable SNR, e.g. -20 dB SNR, the detection rate rises from 0.25 to 0.42. Here again, almost twice as many MBs are detected.

When SNR exceeds 10 dB, the system's contribution is no longer obvious. The detection rate is even slightly worse after denoising than before. This indicates that some MBs are filtered out by the system. However, as the detection algorithm is relatively inexpensive in terms of computing time, it is perfectly feasible to detect on the noisy and on the denoised signal in parallel. In this way, the detection performance of the system corresponds to the maximum detection rate of the two curves.

## 4.3. Performance on a different caliber

To confirm the contribution of this method, shots from a different caliber of ammunition are used for scoring (prediction only). Changing the caliber of ammunition necessarily changes not only the type of weapon used, but also the cartridge, resulting in a different detonation and therefore an acoustically different MB. We can therefore consider this data as a test database for our neural network.

As in section 4.2, Figure 8 shows the performance obtained with the denoising process presented in section 3.2. The evolution of the MB detection rate without vehicle noise as a function of SNR is shown in green. The curve in orange represents these same shots after the addition of moving military vehicle noise, and in blue these signals after being denoised by the proposed system.

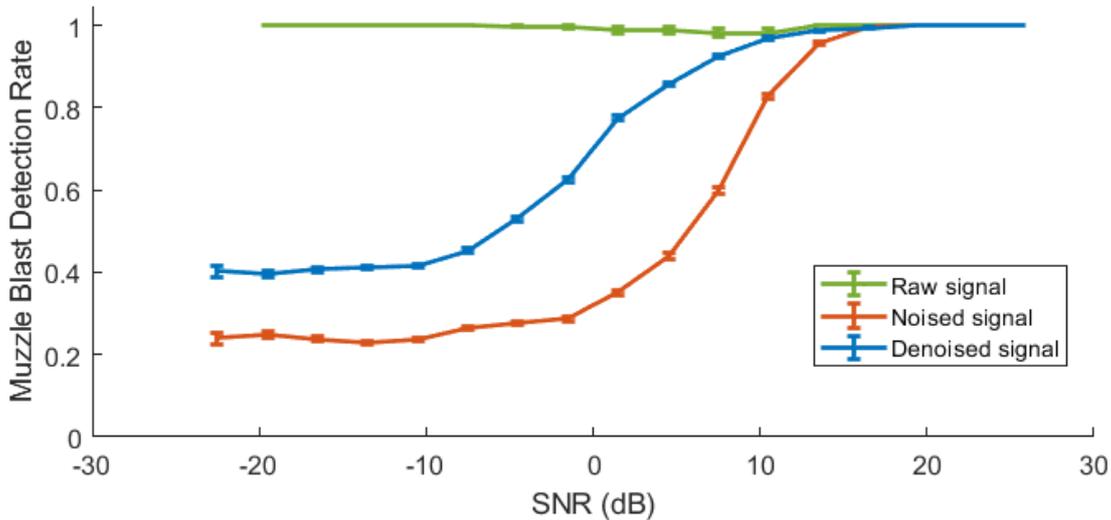

Figure 8Figure 8 : Improved MB detection rate thanks to denoising on test datasets

While detection performance with and without noise is similar to that shown in Figure 7, the performance of the denoising system is slightly lower. This is particularly true in the SNR range [-10 dB; 0 dB]. For example, when the SNR is -5 dB, the detection rate is 0.55 on this test dataset, compared with 0.7 on our validation dataset.

Despite this relative drop in performance, the filtering system still delivers quite acceptable results, thus demonstrating the robustness of the developed model. This was the objective of this test as, for further improvements, the process will be generalized to take advantage, in training, of samples of all calibers.

The most visible contribution of the denoising system is around 0 dB SNR. In this case, the detection rate rises from 0.25 to 0.62. But even when the SNR is -20 dB, the detection rate rises from 0.25 before filtering to 0.4 after filtering. This confirms the significant contribution of the filtering system, even on shots whose nature is different from that on which filtering neural network has been trained.

## CONCLUSION

The objective of this work was to develop a neural network to filter the noise of a moving military vehicle in order to detect the muzzle blast generated by a gunshot. The main constraint was that the chosen architecture had to be algorithmically simple in order to be easily portable on the DSP of Acoem's existing acoustics shots detection hardware platform. The neural network also had to be able to process the signal in real time.

To meet these specifications, an original architecture inspired by autoencoders was proposed. The originality of the work lies in the use of an autoencoder followed by a filtering operation. This filtering operation is translated into a matrix-vector product, so that it can be seen as part of a perceptron-type neural network, and can be optimised by supervised learning at the same time as the autoencoder.

Finally, the training sequence of the neural network thus obtained is split into different sequences to improve performance. In the end, the proposed network makes it possible to increase the muzzle blast detection rate from 0.25 to 0.62 on muzzle blast coming from a caliber of different size from thus used for the learning, with a SNR of 0 dB. These performances may therefore be increased by training the neural network on a dataset with more different types of calibers.